\documentclass[submission,copyright,creativecommons]{eptcs}
\providecommand{\event}{SAIRP 2013}
\usepackage{breakurl} 

\usepackage[T1]{fontenc}
\usepackage{latexsym}
\usepackage{amssymb}

\usepackage[latin1]{inputenc}

\newcommand{\notunlhd}{~/\!\!\!\!\!\!\unlhd}

\newtheorem{theorem}{Theorem}

\title{A Comparison of Well-Quasi Orders on Trees}

\author{Torben Æ.\ Mogensen
\institute{DIKU, University of Copenhagen}
\email{torbenm@diku.dk}}

\def\titlerunning{A Comparison of Well-Quasi Orders on Trees}
\def\authorrunning{T.Æ.\ Mogensen}

\begin{document}

\maketitle

\begin{abstract}
Well-quasi orders such as homeomorphic embedding are commonly used to
ensure termination of program analysis and program transformation, in
particular supercompilation.

We compare eight well-quasi orders on how discriminative they are and
their computational complexity.  The studied well-quasi orders
comprise two very simple examples, two examples from literature on
supercompilation and four new proposed by the author.

We also discuss combining several well-quasi orders to get well-quasi
orders of higher discriminative power.  This adds 19 more well-quasi
orders to the list.

\end{abstract}

\section{Introduction}

A quasi order $(S,\,\leq)$ is a set $S$ with a reflexive and
transitive binary relation $\leq$ on $S\times S$.

A well-quasi order $(S,\,\unlhd)$ is a quasi order such that for any
infinite sequence of elements $s_0, s_1, \ldots$ where $\forall i \in
\mathbb{N}: s_i \in S$, there exists $i,j \in \mathbb{N}$ such that
$i<j$ and $s_i \unlhd s_j$.  We abbreviate ``well-quasi order'' to
WQO.  When the set $S$ is clear from the context, we will sometimes
refer to a WQO only by the relation $\unlhd$.

Well-quasi orders are often used to ensure termination of program
analyses and program transformations (such as partial evaluation and
supercompilation).  The idea is that when a sequence of values is
produced, production of a new value $s_j$ is seen as ``safe'' if there
is no previous $s_i,~i<j$ such that $s_i \unlhd s_j$.  If a value is
found so $s_i \unlhd s_j$, $s_i$ is replaced by a combination of $s_i$
and $s_j$ in a process called \emph{widening} or \emph{generalisation}
and the sequence is recomputed from this point in the hope that the
sequence will be regular (i.e, that it can ``loop back'' to create a
finite cycle).  See \cite{Leuschel:1998,Leuschel:2002} for a detailed
discussion on the use of a particular well-quasi order called
\emph{homeomorphic embedding} for online termination control.

Sequences of trees are often used in program transformation (and
sometimes in program analysis), and since the set of trees over a
finite signature is usually infinite, some form of widening or
generalisation is required to ensure termination of such sequences.

We can compare different well-quasi orders for the same set by
\emph{discriminative power}: A WQO $(S,\, \unlhd_1)$ is more
discriminative than a WQO $(S,\,\unlhd_2)$ if $\forall s_1, s_2 \in S:~
s_1 \unlhd_1 s_2 \Rightarrow s_1 \unlhd_2 s_2$, and it is strictly
more discriminative if, additionally, $\exists s_1, s_2 \in S:~ s_1
{\notunlhd}_1 s_2 \wedge s_1 \unlhd_2 s_2$.

A more discriminative WQO will allow longer sequences before
generalisation or widening is applied, but (being a WQO) will still
avoid infinite sequences.  Since widening or generalisation imply loss
of information, this can give more precise analysis and stronger
program transformations.  On the flip side, longer sequences before
generalisation or widening is applied imply longer running time and
higher memory use of the analysis or transformation, and transformed
programs can also be larger, as fewer parts of the transformed program
are merged.  So there is a trade off involved.  By presenting a range
of WQOs with different discriminative power and computational cost, we
hope to give researchers a basis for choosing WQOs that works well for
their analyses or transformation systems.

We will for simplicity of the presentation assume that all trees are
built from a finite signature: A finite set of constructors with
finite arities.  It is, however, not hard to generalise to the trees
where constructors can have multiple (or even unbounded) arities.  The
size of a tree can be computed in several ways, for example the number
of nodes, the number of edges or the sum of these two.  We will use
the number of nodes (constructor occurrences) as our measure.

\section{Properties of well-quasi orders}

We will review a few properties of WQOs that we will use in
this paper.

\begin{theorem}\label{finWQO}
If $S$ is a finite set, $(S,\,=)$ is a WQO.
\end{theorem}

Proof: Any infinite sequence of elements from a finite set will repeat
elements.  \hfill $\Box$

\begin{theorem}\label{mapWQO}
If $(S_2,\,\unlhd)$ is a WQO and $f$ is a total function from $S_1$ to
$S_2$, then $(S_1,\,\unlhd^f)$ is a WQO, where $\unlhd^f$ is defined by $x
\unlhd^f y$ iff $f(x) \unlhd f(y)$.
\end{theorem}

Proof: If $s_0, s_1, \ldots$ is an infinite sequence of elements from
$S_1$, then $f(s_0), f(s_1), \ldots$ is an infinite sequence of
elements from $S_2$.  Since $(S_2,\,\unlhd)$ is a WQO, there must be
$i<j$ such that $f(s_i) \unlhd f(s_j)$ which is the definition of $s_i
\unlhd^f s_j$.  \hfill $\Box$

\begin{theorem}[Kruskal, 1960]\label{Kruskal}
If $T$ is a set of finite trees, $(T,\,\unlhd_H)$ is a WQO, where $s
\unlhd_H t$ is defined by the rules:

\[\begin{array}{lcl}
c(s_1,\ldots,s_n) \unlhd_H c(t_1,\ldots,t_n) &\Leftarrow&
 s_1 \unlhd_H t_1 \wedge \cdots \wedge  s_n \unlhd_H t_n \\
s \unlhd_H c(t_1,\ldots,t_n)  &\Leftarrow&
\exists 1\leq i\leq n: s \unlhd_H t_i \\
s ~{\notunlhd}_H~ t &,& \textrm{otherwise}
\end{array}\]

\end{theorem}

Basically, $s \unlhd_H t$ if $s$ can be obtained from $t$ by
repeatedly replacing a subtree $x$ of $t$ by one of the subtrees of
$x$.  This ordering is called \emph{homeomorphic embedding}.

Proof: The proof that $(T,\,\unlhd_H)$ is a well-quasi order can be
found in~\cite{Kruskal:1960}.

\begin{theorem}\label{seqWQO}
If $Q = S^*$ is the set of finite sequences over a finite set $S$,
then $(Q,\,\ll)$, where $\ll$ is the subsequence relation, is a WQO.
\end{theorem}

Proof: We map $Q$ to a set of trees $T$ using the mapping $q$:

\[\begin{array}{lcl}
q(\epsilon) & = & \mbox{a leaf node labeled } \epsilon \\
q(aw) &= & \mbox{a node labeled } a \mbox{ with a single child } q(w)
\end{array}\]

\noindent
We note that $w_1 \ll w_2$ iff $q(w_1) \unlhd_H q(w_2)$, so by
Theorems~\ref{Kruskal} and~\ref{mapWQO}, $(Q,\,\ll) =
(Q,\,\unlhd^q_H)$ is a WQO.  \hfill $\Box$

\begin{theorem}\label{multiWQO}
If $B$ is the set of bags (multisets) over a finite set $S$, then
$(B,\,\subseteq)$, where $\subseteq$ is the subset relation on
multisets, is a WQO.
\end{theorem}

Proof: We map multisets to sequences by sorting the elements based on
any total ordering of $S$.  We then note that $b_1 \subseteq b_2$ iff
$sort(b_1) \ll sort(b_2)$, so by Theorems~\ref{seqWQO}
and~\ref{mapWQO}, $(B,\,\subseteq) = (B,\,\ll^{sort})$ is a WQO.  An
alternative proof uses that $(B,\,\subseteq)$ is a \emph{multiset
  extension}~\cite{Wehrman:2006} of $(S,\,=)$. \hfill $\Box$

\begin{theorem}\label{inf-subsequence}
If $\unlhd$ is a well-quasi order, then
any infinite sequence $s_0, s_1, \ldots$ contains an infinite
increasing subsequence $s_{i_0} \unlhd s_{i_1} \unlhd \ldots$
\end{theorem}

Proof: Assume that the set $M = \{i \,|\, \not\exists k>i: s_i \unlhd
s_k\}$ is infinite.  As $\unlhd$ is a well-quasi order, there must be
$i,j \in M$ such that $i<j$ and $s_i \unlhd s_j$.  But this
contradicts the definition of $M$.  Hence, $M$ is finite and has a
maximal element $m$.  Now assume that the longest increasing
subsequence starting from $i_{m{+}1}$ is finite.  If so, it has a
maximal element $s_p$, such that there are no $q>p:~s_p\unlhd s_q$.
But that would make $p\in M$, which, because $p>m$, contradicts the
fact that $m$ is maximal in $M$. Hence, we have an infinite increasing
subsequence starting from $i_{m{+}1}$. \hfill $\Box$

\begin{theorem}\label{combWQO}
Given two WQOs $(S,\,\unlhd_1)$ and $(S,\,\unlhd_2)$, then $(S,~
{\unlhd_1\cap\unlhd_2})$ defined by $x\, (\unlhd_1\cap\unlhd_2)\, y
\Leftrightarrow x \unlhd_1 y \wedge x \unlhd_2 y$ is a WQO.
\end{theorem}

Proof: Any infinite sequence $s_0, s_1, \ldots$ will due to
Theorem~\ref{inf-subsequence} have an infinite increasing subsequence
$s_{i_0} \unlhd_1 s_{i_1} \unlhd_1 \ldots$, which (because $\unlhd_2$
is a WQO) will have a pair $s_{i_j} \unlhd_2 s_{i_k}$, where $j<k$.
It follows that $s_{i_j}\, (\unlhd_1\cap\unlhd_2)\, s_{i_k}$.  \hfill
$\Box$

\section{A selection of well-quasi orders on trees}\label{WQO-list}

In the following, we will describe six WQOs on trees with finite
signatures.  We divide these WQOs into groups based on whether they
are defined directly on the trees or by mapping trees to another WQO
using Theorem~\ref{mapWQO}.

We use $T$ to denote an otherwise unspecified set of trees over a
finite signature $\Sigma$.  We will also use $\Sigma$ to denote the
set (alphabet) of constructor symbols in $\Sigma$.  The context should
make it clear which meaning is used.

\subsection{Well-quasi orders defined directly on trees}

\paragraph{$\unlhd_S$:}

A simple WQO for trees is based on comparison of size: For any two
trees $t_1, t_2$, we define $t_1 \unlhd_S t_2$ iff $t_1 = t_2$ or
$|t_1|<|t_2|$, where $|t|$ is the size of the tree $t$.  It is clear
that an infinite sequence of trees must either repeat specific trees
or increase the size of trees: There are only finitely many different
trees of a given size.  Hence, $(T,\unlhd_S)$ is a WQO.

\paragraph{$\unlhd_H$:}

A WQO that is often used for controlling termination of program
transformation is \emph{homeomorphic embedding} $(T,\,\unlhd_H)$ as
defined as in Theorem~\ref{Kruskal}.

Homeomorphic embedding has been used for termination proofs for
term-rewriting systems~\cite{Dershowitz:1979}.  Using homeomorphic
embedding to control termination of supercompilation was first
proposed in~\cite{Sorensen:1995}.

\subsection{Well-quasi orders defined by mapping to sets}

\paragraph{$\unlhd_Z$:}

If $S$ is a finite set, Theorem~\ref{finWQO} gives that $(S,\,=)$ is a
WQO.  Since $T$ uses a finite signature $\Sigma$, $(2^\Sigma,\,=)$ is
a WQO, where $2^\Sigma$ is the set of subsets of $\Sigma$.

We define $(T,\,\unlhd_Z)$ by the function $f$ from $T$ to $2^\Sigma$
that maps a tree $t$ to the set of constructors used in $t$, so $s
\unlhd_Z t$ iff $s$ and $t$ use the same set of constructors.
Since$(2^\Sigma,\,=)$ is a WQO, $(T,\,\unlhd_Z) = (T,\,=^f)$ is by
Theorem~\ref{mapWQO} also a WQO.

\paragraph{$\unlhd_Y$:}

We propose a variant of $\unlhd_Z$ by mapping a tree to a set of
constructors using a different mapping: $w$ maps a tree $t$ to the set
of constructors used \emph{at least twice} in $t$.  By the same
reasoning as above, $(T,\,\unlhd_Y) = (T,\,=^w)$ is WQO.  Obviously,
this can be generalised to sets of constructors that are used at least
3, 4 or more times.

$\unlhd_Y$ is appropriate for supercompilation and related
transformations, as an infinite sequence of trees usually copies some
part of an earlier initial tree in more and more copies.  Such copying
will, eventually, create a tree with the same set of constructors that
are used at least twice as an earlier tree.  But $\unlhd_Y$ will not
stop temporary growth that does not preserve this set, such as
replacing a subtree $c(a,a,b)$ by $c(a,b,b)$.  Obviously, larger trees
will trigger more false positives, so $\unlhd_Y$ works best in
combination with other WQOs.

\subsection{Well-quasi orders defined by mapping to multisets}

For these WQOs we map a tree to the multiset (bag) of its
constructors.  More precisely, we use a function $g$ from $T$ to the
set $B = \mathbb{N}^\Sigma$ of multisets over $\Sigma$ defined in the
following way:

\[\begin{array}{lcl}
g(c) &=& \{c\}\\
g(c(s_1,\ldots,s_n)) &=& \{c\} \cup \bigcup_{i=1}^n g(s_i)
\end{array}\]

\noindent
where $\cup$ is union on multisets.

\paragraph{$\unlhd_B$:}

We propose a new WQO $(T,\,\unlhd_B)$ that directly uses $g$: $s
\unlhd_B t \Leftrightarrow g(s) \subseteq g(t)$.

$(B,\,\subseteq)$ is a WQO due to Theorem~\ref{multiWQO}, so
$(T,\,\unlhd_B) = (T,\,\subseteq^g)$ is a WQO by Theorem~\ref{mapWQO}.

We deem $\unlhd_B$ appropriate for supercompilation and related
transformations because an infinite sequence usually involves copying
nodes in the tree more and more times, which would make the multiset
of constructors in a new tree a superset of those in a previous tree.

\paragraph{$\unlhd_M$:}

Another WQO that has been used for
supercompilation~\cite{Mitchell:2010} is also based on multisets over
$\Sigma$.

Given two multisets $b_1, b_2 \in B$, we define $b_1 \leq b_2
\Leftrightarrow b_1=b_2 \vee set(b_1) = set(b_2) \wedge |b_1|<|b_2|$,
where $set(b)$ is the set of different elements in $b$ and $|b|$ is
the total number of elements in $b$.\footnote{Mitchell uses a negated
  form of this relation and negates the test for termination
  detection.}

Since there are only finitely many different sets over a finite
alphabet, any infinite sequence must have infinitely many bags with
the same underlying sets.  Since there are only finitely many bags
with the same size, this means that, in any infinite sequence, the
size of the bags must increase.  Hence, $(B,\,\leq)$ is a WQO.

We define $(T,\,\unlhd_M)$ to be the quasi order on trees derived from
$(B,\,\leq)$ by the mapping $g$.  In other words, $s \unlhd_M t
\Leftrightarrow g(s) \leq g(t)$.  By Theorem~\ref{mapWQO},
$(T,\,\unlhd_M) = (T,\,\leq^g)$ is a WQO.  Basically, $\unlhd_M$
refines $\unlhd_S$ by making trees incomparable if they have different
underlying sets.

In~\cite{Mitchell:2010}, subexpressions of the original program are
named, and new expressions inherit the names of their progenitors.  So
any expression will be assigned a multiset of names.  Since names of
new expressions are based on where in the original program their
progenitor expressions occur, two identical expression trees can have
different multisets of names.  While adding names to nodes in the
trees adds information that a plain expression tree does not have, the
mapping from trees to multisets is basically the same as $g$, except
that a node with multiple names is mapped to a multiset of all these
names instead of to a multiset with just one name.

\subsection{Well-quasi orders defined by mapping to strings}

\paragraph{$\unlhd_P$:}

We map trees into strings over a finite alphabet and compare these
strings.  We map a tree $t$ to a string $w$ by a mapping $Pre$ from
$T$ to $\Sigma^*$, i.e, the set of finite strings over the alphabet
$\Sigma$.  $Pre(t)$ is the string of constructors in $t$ in
preorder-traversal order:

\[\begin{array}{lcl}
Pre(c) &=& c \\
Pre(c(t_1,\ldots,t_n) &=& c\, Pre(t_1) \cdots Pre(t_n)
\end{array}\]

\noindent
Note that, since constructors have fixed arities, $Pre$ is injective:
Different trees map to different strings.

We compare the strings using the subsequence order: $w_1 \ll w_2$ if
$w_1$ is a subsequence of $w_2$.  By Theorem~\ref{seqWQO},
$(\Sigma^*,\,\ll)$ is a WQO.

We now define the quasi order $(T,\,\unlhd_P)$ by $t_1 \unlhd_P t_2
\Leftrightarrow Pre(t_1) \ll Pre(t_2)$.  Theorem~\ref{mapWQO} gives
that $(T,\,\unlhd_P) = (T,\,\ll^{Pre})$ is a WQO.

$\unlhd_P$ is appropriate for supercompilation because it approximates
$\unlhd_H$, which has proven successful for controlling termination in
supercompilation, but (as we shall see) $\unlhd_P$ is less costly to
compute.

\paragraph{$\unlhd_E$:}

We can refine $\unlhd_P$ by mapping trees to strings that contain more
information about the structure of the tree: In a preorder traversal,
a constructor node is visited only once, before its children.  But we
can add additional visits of the node between and after traversing its
children.  This is sometimes called an Euler-tour traversal.  We
define this through a modified traversal function:

\[\begin{array}{lcl}
Eul(c) &=& c \\
Eul(c(t_1,\ldots,t_n) &=& c_0\, Eul(t_1)\,c_1 \cdots Eul(t_n)\,c_n
\end{array}\]

\noindent
where $c_i$ is a symbol that marks that $i$ children of a constructor
node labelled $c$ have been traversed.

Theorems~\ref{seqWQO} and~\ref{mapWQO} give that $(T,\,\unlhd_E) =
(T,\,\ll^{Eul})$ is a WQO.

$\unlhd_E$ is appropriate for supercompilation for the same reason
that $\unlhd_P$ is.

\section{Comparing well-quasi orders by discriminative power}

We recall that a WQO $\unlhd_1$ is more discriminative than a WQO
$\unlhd_2$ if $\forall s_1, s_2 \in S:~ s_1 \unlhd_1 s_2 \Rightarrow
s_1 \unlhd_2 s_2$, and it is strictly more discriminative if,
additionally, $\exists s_1, s_2 \in S:~ s_1 {\notunlhd_1} s_2 \wedge
s_1 \unlhd_2 s_2$.

We will use the following trees in our discussion about the relative
discriminative power of WQOs:

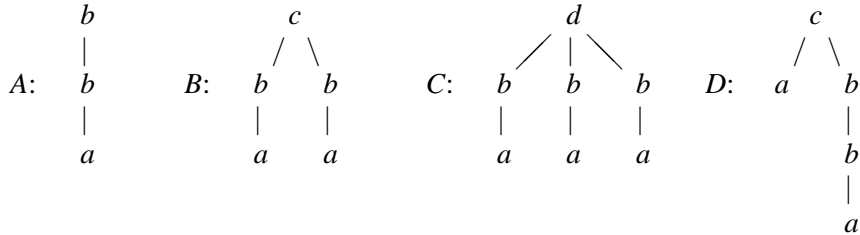
\begin{center}
\setlength{\unitlength}{0.6em}
\begin{picture}(50,14)(0,-3)
\put(0,5){$A$:}
\put(4,9){$b$}
\put(4,5){$b$}
\put(4,1){$a$}
\put(4,3){$|$}
\put(4,7){$|$}

\put(10,5){$B$:}
\put(16,9){$c$}
\put(14,5){$b$}
\put(14,1){$a$}
\put(18,5){$b$}
\put(18,1){$a$}
\put(14,3){$|$}
\put(18,3){$|$}
\put(15,7){$/$}
\put(17,7){$\backslash$}

\put(24,5){$C$:}
\put(32,9){$d$}
\put(28,5){$b$}
\put(28,1){$a$}
\put(32,5){$b$}
\put(32,1){$a$}
\put(36,5){$b$}
\put(36,1){$a$}
\put(28,3){$|$}
\put(32,3){$|$}
\put(32,7){$|$}
\put(36,3){$|$}
\put(29.1,6.4){\line(1,1){2}}
\put(35.2,6.4){\line(-1,1){2}}

\put(40,5){$D$:}
\put(46,9){$c$}
\put(44,5){$a$}
\put(48,5){$b$}
\put(48,1){$b$}
\put(48,-3){$a$}
\put(48,-1){$|$}
\put(48,3){$|$}
\put(45,7){$/$}
\put(47,7){$\backslash$}

\end{picture}
\end{center}

\vspace{2ex}

\noindent
It is not hard to see that $t_1 \unlhd_H t_2 \Rightarrow Eul(t_1) \ll
Eul(t_2) \Leftrightarrow t_1 \unlhd_E t_2$, so $\unlhd_H$ is more
discriminative than $\unlhd_E$.  It is also \emph{strictly} more
discriminative:

\[Eul(A) = b_0b_0ab_1b_1 ~\ll~ d_0b_0ab_1d_1b_0ab_1d_2b_0ab_1d_3 = Eul(C)\]

\noindent
so $A \unlhd_E C$, but it is clear from inspection of the trees that
$A ~{\notunlhd}_H~ C$.

It is also clear that $\unlhd_E$ is more discriminative than
$\unlhd_P$: If $Eul(t_1) \ll Eul(t_2)$ then, clearly, $Pre(t_1)
\ll Pre(t_2)$.  It is also strictly more discriminative: $Eul(A) =
b_0b_0ab_1b_1 ~{\not\ll} ~ c_0b_0ab_1c_1b_0ab_1c_2 = Eul(B)$, but
$Pre(A) = bba \ll cbabab = Pre(B)$, so $A ~{\notunlhd}_E~ B$
while $A \unlhd_P B$.

It is easy to see that $\unlhd_P$ is more discriminative than
$\unlhd_B$: If $Pre(t_1)\ll Pre(t_1)$, then $bag(Pre(t_1))\subseteq
bag(Pre(t_1))$, where $bag$ maps a string to a bag by ignoring the
order of elements.  It is also strictly more discriminative, as $B
~{\notunlhd}_P~ D$ but $B \unlhd_B D$, since the trees $B$ and $D$
map to the same bag of constructors $\{a,a,b,b,c\}$.

It is, however, not the case that $\unlhd_B$ is more discriminative
than $\unlhd_M$.  We see this by observing that $(B,\,\subseteq)$ and
$(B,\,\leq)$ have incomparable discriminative power:

On the one hand, $\{a,b,b\} \leq \{a,a,a,b\}$, since their underlying
sets are the same and $|\{a,b,b\}|<|\{a,a,a,b\}|$, but on the other
hand $\{a,b,b\} ~{\not\subseteq}~ \{a,a,a,b\}$.  Also, $\{a\}
\subseteq \{a,b\}$ but $\{a\}~ {\not\leq}~ \{a,b\}$, because the
underlying sets are different.  In fact, because $\unlhd_M$ makes
trees with different underlying sets of constructors incomparable, its
discriminative power is also incomparable to $\unlhd_P$ and
$\unlhd_H$.

It is trivial to see that $\unlhd_M$ is strictly more discriminative
than $\unlhd_S$, as $\unlhd_M$ is effectively size comparison
restricted to trees with the same underlying set of constructors.

$\unlhd_B$ has incomparable discriminative power to $\unlhd_S$ also:
Two different trees of the same size are incomparable by $\unlhd_S$,
but if they are built from the same multiset of constructors, they are
comparable by $\unlhd_B$.

But $\unlhd_S$ has strictly less discriminative power than $\unlhd_P$
(and, hence, also $\unlhd_E$ and $\unlhd_H$): Since constructors have
fixed arity, two trees have the same preorder traversal strings only
if they are identical.  Two same-size trees will have same-length
preorder-traversal strings, and same-length strings can be in the
subsequence relation only if they are equal.  So two non-identical
same-size trees will be incomparable by both $\unlhd_P$ and
$\unlhd_S$.  But where trees of different size are always comparable
by $\unlhd_S$, they need not be comparable by $\unlhd_P$.

$\unlhd_Z$ is clearly strictly less discriminative than $\unlhd_M$, as
both consider the underlying set but $\unlhd_Z$ does not consider the
size of the trees (or, equivalently, the size of the underlying
multisets).  The discriminative powers of $\unlhd_Z$ and $\unlhd_S$
are incomparable, as are the discriminative powers of $\unlhd_Z$ and
$\unlhd_B$.

$\unlhd_Z$ and $\unlhd_Y$, though related, are of incomparable
discriminative power: The trees $B$ and $C$ are comparable by
$\unlhd_Y$, as the constructors $a$ and $b$ (and none other) occur
twice or more in both trees, but the underlying sets are different so
they are not comparable by $\unlhd_Z$.  On the other hand $D$ and a
tree similar to $D$ but with only one $b$ constructor will be
comparable by $\unlhd_Z$ but not by $\unlhd_Y$.  $\unlhd_Y$ is, in
fact, incomparable with all the other WQOs described in
Section~\ref{WQO-list}.

This gives us the following hierarchy of discriminative power:

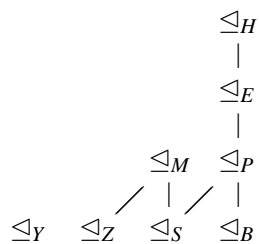
\begin{center}
\setlength{\unitlength}{0.6em}
\begin{picture}(13,15)(-6,0)
\put(-0.9,2.3){\line(1,1){1.6}}
\put(-7,1){$\unlhd_Y$}
\put(-3,1){$\unlhd_Z$}
\put(1,1){$\unlhd_S$}
\put(1,5){$\unlhd_M$}
\put(5,1){$\unlhd_B$}
\put(5,5){$\unlhd_P$}
\put(5,9){$\unlhd_E$}
\put(5,13){$\unlhd_H$}
\put(2,3){|}
\put(6,3){|}
\put(6,7){|}
\put(6,11){|}
\put(3.1,2.3){\line(1,1){1.6}}
\end{picture}
\end{center}

\subsection{Combining well-quasi orders}

We can combine two or more incomparable WQOs by intersection to get a
WQO of more discriminative power.

For example, we can define a WQO~$(T,\,\unlhd_{SB}) =
(T,\,{\unlhd_S\cap\unlhd_B})$.\footnote{When we subscript $\unlhd$
  with several letters, this denotes the intersection of the orderings
  defined by $\unlhd$ subscripted with each individual letter.} In
other words, $s \unlhd_{SB} t \Leftrightarrow s \unlhd_{S} t \wedge s
\unlhd_{B} t$.  Since $\unlhd_{S}$ and $\unlhd_{B}$ have incomparable
discriminative power, it clear that $\unlhd_{SB}$ is strictly more
discriminative than both $\unlhd_{S}$ and $\unlhd_{B}$.  $\unlhd_{SB}$
has strictly less discriminative power than $\unlhd_P$: Consider a
tree $D'$ similar to $D$ but with one extra $b$ node in the right
branch.  It is clear that $B \unlhd_{S} D'$, since $D'$ is bigger than
$B$ and $B \unlhd_{B} D'$ since the bag of constructors in $B$ is a
subset of the bag of constructors in $D'$.  But $B\, {\notunlhd}_P\,
D'$ since $Pre(B) \not\ll Pre(D')$.

From its definition, it is easy to see that $(T,\,\unlhd_M) = (T,\,
{\unlhd_{ZS}})$, where $\unlhd_{ZS} = (\unlhd_Z \cap\unlhd_S)$.  All
other combinations of uncomparable WQOs yield new WQOs.  For example,
as we saw above, $\unlhd_{SB}$ is less discriminative than
$\unlhd_{P}$ but more discriminative than both $\unlhd_{S}$ and
$\unlhd_{B}$.

Adding all combinations of incomparable WQOs introduced in
Section~\ref{WQO-list} gives a hierarchy of WQOs which is induced from
the partial order shown in the hierarchy above.  The only non-obvious
results of the combinations are that $\unlhd_{SB}$ is strictly below
$\unlhd_{P}$ and that $\unlhd_M = \unlhd_{ZS}$, which implies, for
example, $\unlhd_{MP} = \unlhd_{ZSP} = \unlhd_{ZP}$.  The complete
hierarchy is too complicated to show graphically, so we show below the
hierarchy of the 13 single and combined WQOs that do not involve
$\unlhd_{Y}$.  For each of these, there is a more discriminative WQO
obtained by combining it with $\unlhd_{Y}$.  Adding also $\unlhd_{Y}$
itself brings the total count up to 27.

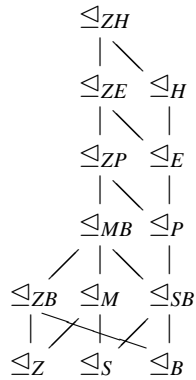
\begin{center}
\setlength{\unitlength}{0.6em}
\begin{picture}(13,23)(-6,0)
\put(-3,1){$\unlhd_Z$}
\put(1,1){$\unlhd_S$}
\put(-3,5){$\unlhd_{ZB}$}
\put(1,5){$\unlhd_{M}$}
\put(1,9){$\unlhd_{MB}$}
\put(1,13){$\unlhd_{ZP}$}
\put(1,17){$\unlhd_{ZE}$}
\put(1,21){$\unlhd_{ZH}$}
\put(5,1){$\unlhd_B$}
\put(5,5){$\unlhd_{SB}$}
\put(5,9){$\unlhd_P$}
\put(5,13){$\unlhd_E$}
\put(5,17){$\unlhd_H$}
\put(-2,3){$|$}
\put(2,3){$|$}
\put(2,7){$|$}
\put(2,11){$|$}
\put(2,15){$|$}
\put(2,19){$|$}
\put(6,3){$|$}
\put(6,7){$|$}
\put(6,11){$|$}
\put(6,15){$|$}
\put(4.9,2.3){\line(-3,1){6}}
\put(-0.7,2.3){\line(1,1){1.7}}
\put(3.3,2.3){\line(1,1){1.7}}
\put(-0.7,6.3){\line(1,1){1.7}}
\put(4.7,6.3){\line(-1,1){1.7}}
\put(4.7,10.3){\line(-1,1){1.7}}
\put(4.7,14.3){\line(-1,1){1.7}}
\put(4.7,18.3){\line(-1,1){1.7}}
\end{picture}
\end{center}

\noindent
This hierarchy says only \emph{whether} one of the listed WQOs is more
discriminative than another, but not \emph{how much} more
discriminative it is.  There is no definitive measure for this, but we
have made an approximate measure by generating 400 random, different
trees using a signature with one nullary constructor \texttt{a}, one
unary constructor \texttt{b}, one binary constructor \texttt{c} and
one ternary constructor \texttt{d}, using the following probabilities
for the different constructors:

\[\begin{array}{c@{\quad}r}
 \texttt{a} & 50\% \\
 \texttt{b} & 20\% \\
 \texttt{c} & 15\% \\
 \texttt{d} & 15\%
\end{array}\]

\noindent
The table below shows for each of the discussed WQOs and all relevant
combinations of these the number of pairs $(t_1,t_2)$ (out of
$160\,000$) where $t_1\unlhd t_2$.

\[\begin{array}[t]{l@{\quad}r}
\unlhd_{YZH} & 709 \\
\unlhd_{YH} & 899 \\
\unlhd_{YZE} & 3659 \\
\unlhd_{YE} & 3875 \\
\unlhd_{ZH} & 5483 \\
\unlhd_{YZP} & 6641 \\
\unlhd_{YP} & 6999 \\
\unlhd_{ZE} & 12671 \\
\unlhd_{YMB} = \unlhd_{YZSB} & 17632 \\
\unlhd_H & 18587 \\
\unlhd_{YSB} & 18662 \\
\unlhd_{YZB} & 19208 \\
\unlhd_{YB} & 19238 \\
\end{array}\quad\quad\quad\quad\quad\quad
\begin{array}[t]{l@{\quad}r}
\unlhd_{ZP} & 19355 \\
\unlhd_{YM} = \unlhd_{YZS} & 22868 \\
\unlhd_{YS} & 24579 \\
\unlhd_E &  29252 \\
\unlhd_{MB} = \unlhd_{ZSB} & 37551 \\
\unlhd_{ZB} & 38127 \\
\unlhd_P &  39168 \\
\unlhd_{YZ} & 44384 \\
\unlhd_Y & 47642 \\
\unlhd_M = \unlhd_{ZS} & 47915 \\
\unlhd_{SB} & 61480 \\
\unlhd_B & 62056 \\
\unlhd_S & 78582 \\
\unlhd_Z & 93870 \\
\end{array}\]

\noindent
Though this is a very rough comparison, as the random trees are hardly
typical of what you see in program transformation and analysis, it
shows a considerable difference in discriminative power between the
different WQOs and also that combining two incomparable WQOs can yield
a WQO with significantly higher discriminative power.

\section{Comparing well-quasi orders by computational complexity}

The typical scenario is that a new element in a sequence is compared
to all previous elements in the sequence.  This means that the total
number of comparisons required for an $n$-long sequence is
$n(n{-}1)/2$ or O($n^2$).

So a way to reduce the overall complexity is to, if possible, split a
comparison $s \unlhd t$ into two steps: First computing values
$f(s),\,f(t)$ and then comparing these using a computationally simpler
ordering (essentially using Theorem~\ref{mapWQO}).  $f(s_i)$ is
computed only once per element $s_i$ in the list, so even a small
saving in comparing the results will give an overall saving, even if
precomputing $f(s_i)$ is relatively expensive.  We will primarily
focus on the accumulated cost of building a sequence and comparing
each new element to all previous elements, using any relevant
precomputation on each element.  We also exploit the fact that we stop
building the sequence as soon as we add an element $t$ such that
there is a $s_i \unlhd t$ in the sequence already.

As a simple example, consider $\unlhd_S$, which is defined by $s
\unlhd_S t \Leftrightarrow s = t \vee |s|<|t|$.  Clearly, the
comparison of the sizes (once these are computed) is very fast, but
time for computation of the sizes is proportional to the sizes
themselves.  If all $n$ trees have sizes close to $S$, an
implementation that computes the size at every comparison would need
O($n^2\cdot S$) time, but if the sizes are computed only once per
tree, the total cost of size comparisons is only O($n\cdot S + n^2$),
assuming unit cost of integer comparison.  On top of the size
comparison, we must compare a new element for equality to all previous
elements.  Though comparing for equality is in the worst case
proportional to the size of the trees, it can in practice be made
(near) constant time by using hashing, requiring an O($S$)
precomputation per tree for computing the hash.  So, assuming
effective hashing, the identity comparisons do not add to the overall
asymptotic complexity.  Additionally, since we stop adding elements to
a sequence once we find an element that compares to a previous
element, the sizes of elements in a sequence are non-increasing.  This
means that we only need to compare the size of a new element $t$ with
that of the last previously added element $s_i$.  And comparing $t$
for identity with each previous element can be avoided by using the
hash code to look up in a table that indicates whether a tree with the
same hash code has been seen before.  This reduces the cost (again
assuming effective hashing) to O($S$) for each new added element, so
the total cost for an $n$-long sequence is O($n \cdot S$).

For $\unlhd_Z$, we can for each tree precompute its set of
constructors.  The sets can be hashed or (if $\Sigma$ is small)
represented by a short bit vector, so we do an O($S$) precomputation
per tree to compute a hash value or a bit vector.  Much like above, we
can use the hash value or bit vector as a key into a table that
indicates whether we have seen the set before, so again the total cost
O($n\cdot S$).  The same analysis applies to  $\unlhd_Y$.

$\unlhd_B$ also maps trees to multisets, which can be precomputed.  A
multiset can be represented as a vector of the number of occurrences
of each constructor, and the comparison of two such vectors is
proportional to their size, which is the size $|\Sigma|$ of the
alphabet of constructors $\Sigma$.  Precomputation is, again, linear
in the size of the trees, so the total cost is O($n\cdot S +
|\Sigma|\cdot n^2$).

$\unlhd_P$ maps trees to strings and then compares these using the
subsequence order.  Mapping a tree of size $S$ to a string is O($S$)
and produces a string of size $S$.  Deciding the subsequence relation
for two strings $v$ and $w$ is O($|v|+|w|$).  Subsequence tests are
done a total of O($n^2$) times, so the total cost is O($n^2\cdot S$),
which is significantly more than for the previous WQOs.  $\unlhd_E$
has the same asymptotic cost as $\unlhd_P$, but with a higher constant
factor (about twice as high), as the generated strings are roughly
twice as long.

But this is still far less than the cost of $\unlhd_H$.  At the time
of writing, the fastest known method~\cite{Bille:2011} for deciding
$t_1 \unlhd_H t_2$ is O($|t_1|\cdot|t_2| / \log(|t_2|) + |t_2|\cdot
\log(|t_2|)$), which makes the total cost for a sequence of $n$
size-$S$ trees O($n^2 \cdot S^2 / \log(S)$), since the O($S \cdot
\log(S)$) component of the cost is asymptotically dominated by O($n^2
\cdot S^2 / \log(S)$).  There is no obvious precomputation that can
reduce this time.

The table below summarises the costs both for pairwise comparison and
for comparing each element in a sequence of $n$ elements of size $S$
to all previous elements in the sequence.

\[\begin{array}{c@{\quad}c@{\quad}c}
\textrm{WQO} & \textrm{$s \unlhd t$}
 & \textrm{sequence of $n$ trees of size $S$} \\\hline

\unlhd_Z & \textrm{O}(|s|+|t|) & \textrm{O}(n \cdot S) \\

\unlhd_Y & \textrm{O}(|s|+|t|) & \textrm{O}(n \cdot S) \\

\unlhd_S & \textrm{O}(|s|+|t|) & \textrm{O}(n \cdot S) \\

\unlhd_B & \textrm{O}(|s|+|t|)
 & \textrm{O}(n \cdot S + |\Sigma|\cdot n^2) \\

\unlhd_P & \textrm{O}(|s|+|t|) & \textrm{O}(n^2 \cdot S) \\

\unlhd_E & \textrm{O}(|s|+|t|) & \textrm{O}(n^2 \cdot S) \\

\unlhd_H & \textrm{O}(|s|\cdot|t| / \log(|t|) + |t|\cdot \log(|t|))
 & \textrm{O}(n^2 \cdot S^2 / \log(S))
\end{array}\]

\noindent
While many of the WQOs have the same complexity {O}$(|s|+|t|)$ for
pairwise comparison of trees, the accumulated complexity of
constructing and checking a sequence differs considerably for several
of these.

\subsection{Cost of combined WQOs}

With a few exceptions, combined WQOs have asymptotic costs that are
the sums of the asymptotic costs of their components, as you can test
each component WQO independently of the others.

The exceptions are combinations of $\unlhd_S$ and other WQOs as we, to
get the cost of testing a sequence with $\unlhd_S$ down to O($n \cdot
S$), exploited that the sizes of trees in a sequence using $\unlhd_S$
as indicator would stop when the size of the new tree is larger than
the immediately previous, so the sizes of trees in the sequence would
be non-increasing.  But when combining $\unlhd_S$ with another WQO
$\unlhd'$, the sizes of trees in the sequence are not necessarily
non-increasing, as the size of a new element $t$ can be larger than
the size of a previous element $s_i$ as long as $s_i\, {\notunlhd}'\,
t$.  So we will generally have to compare $t$ with all previous
elements $s_i$.  We can still precompute the sizes of the trees and
compute hash codes for quick equality testing, but unless $\unlhd'$
allows partitioning the sequence into subsequences of non-increasing
size, we will need to compare the size of $t$ to the sizes of all
previous elements in the sequence.  So without knowing more about
$\unlhd'$, the part of the total cost needed for testing a sequence
with $\unlhd_S$ is O($n \cdot S + n^2$).  So, generally, the cost of
combining $\unlhd_S$ with a WQO $\unlhd'$ is the sum of O($n \cdot S +
n^2$) and the cost for $\unlhd'$.  This can, however be avoided for
some instances of $\unlhd'$, such as $\unlhd_Y$ and $\unlhd_Z$, as we
shall see below.

If $f$ is a function from $T$ to a finite set $F$ (such as
$2^\Sigma$), then testing a sequence for a combination of a WQO of the
form $(T,\,=^f)$ and another WQO $\unlhd'$ can be done by partitioning
the sequence $s_i$ (as it is built) by different values of $f(s_i)$.
Using hashing on the values of $f(s_i)$, finding the right partition
for a new element $t$ can be done in O($|t|$) time (to perform the
hashing) and a constant-time lookup.  The new element is then compared
to the other elements in the partition using $\unlhd'$ only, so the
total cost of adding an element $t$ to the sequence is O($|t|$) plus
the cost using $\unlhd'$ of adding $t$ to a sequence.  For example,
combining $\unlhd_Z$ and $\unlhd_S$ can be done by partitioning the
sequence by different sets of constructors.  Each partition will be a
subsequence of the original with non-increasing sizes, so the fast
method for testing a sequence for $\unlhd_S$ can be used for each
subsequence, which brings the total cost for building and testing a
sequence with $\unlhd_{ZS} = \unlhd_{M}$ down to O($n \cdot S$).  The
same analysis applies to combining $\unlhd_Y$ with $\unlhd_S$.

All the other WQOs presented in Section~\ref{WQO-list} require
comparison of (some precomputed value of) the new tree $t$ to all
previous trees $s_i$, so the cost of combining two of these is a
simple addition of the costs of the components.  We can, however,
first test a new element $t$ against previous elements $s_i$ using the
cheaper of the two WQOs $\unlhd_1$, and only when that finds an $s_i
\unlhd_1 t$ check if $s_i \unlhd_2 t$, where $\unlhd_2$ is the more
costly of the two.

Similarly, less costly measures can be used to approximate more costly
measures: For example, when comparing with $\unlhd_{YH}$, we can first
test (fairly cheaply) if $s_i\, \unlhd_{YS}\, t$, and if this is true
try $s_i\, \unlhd_E\, t$, and only if this is also true perform the
expensive test $s_i\, \unlhd_H\, t$.  This way, we get the full
discriminative power of $\unlhd_{YH}$ but we will in many (probably
most) cases be able to avoid the full cost.

\section{Conclusion}

We have compared a number of well-quasi orders (WQOs) on trees for
discriminative power and computational cost.  The selection of WQOs
include very simple WQOs ($\unlhd_Z$, $\unlhd_S$), several WQOs from
the literature of supercompilation ($\unlhd_M$, $\unlhd_H$) as well as
some proposed by the author ($\unlhd_Y$, $\unlhd_B$, $\unlhd_P$,
$\unlhd_E$), that to the author's knowledge have not been used for
termination of program analysis and transformation.

We have also looked at combining two or more WQOs of incomparable
discriminative power to get a more discriminative WQO.  This adds 19
more distinct WQOs ($\unlhd_{YZ}$, $\unlhd_{ZH}$, $\unlhd_{YH}$,
$\unlhd_{YZH}$, $\unlhd_{ZE}$, $\unlhd_{YE}$, $\unlhd_{YZE}$,
$\unlhd_{ZP}$, $\unlhd_{YP}$, $\unlhd_{YZP}$, $\unlhd_{ZB}$,
$\unlhd_{YB}$, $\unlhd_{YZB}$, $\unlhd_{SB}$, $\unlhd_{MB}$,
$\unlhd_{YSB}$, $\unlhd_{YMB}$, $\unlhd_{YS}$, $\unlhd_{YM}$).

We observe that, in a typical scenario, we do not just compare a pair
of trees, but compare each new element of a sequence to all previous
elements.  In such a scenario, time can be saved by precomputing some
values for each tree as it is added to the sequence, and then using
these values for the WQO comparison.  This can for many WQOs
dramatically reduce the overall cost.

While higher discriminative power gives more precision, there is a
higher cost not only from computing the more complex WQO, but also
because sequences of trees get longer before they are
generalised/widened and folded back, so analyses and transformations
can take longer (and transformed programs can be bigger). The choice
of which WQO to use should consider all of the above.

When using a combined WQO, it can be cheaper to first compare with the
cheapest component WQO and only if that succeeds, compare with the
more expensive component WQO.  When combining with $\unlhd_Z$ and
$\unlhd_Y$, these can be used to partition the sequence.  This idea
can also be applied with comparable WQOs: If a cheaply computable WQO
approximates a more expensive WQO, we can compute the cheap WQO and
only if that succeeds compute the expensive WQO.  In neither case is
the worst-case cost lowered (and it can be somewhat increased), but
the average cost can be significantly lower.

From the (admittedly naive) statistic tests, it seems combining a WQO
$\unlhd'$ with $\unlhd_Z$ or (in particular) $\unlhd_Y$ gives
significantly higher discriminative power than $\unlhd'$ alone, but
combining with both $\unlhd_Z$ and $\unlhd_Y$ gives only little extra
power over combining with $\unlhd_Y$ alone.

\section{Acknowledgements}

The author would like to thank the organisers of SAIRP for inviting
him to submit a paper.  He also thanks the anonymous reviewers, whose
suggestions greatly improved the quality of the paper.

The author would also like to thank Dave Schmidt for being an
inspirational scientist and a generally fun guy to be around.  The
halo over his head in the photo on his personal home
page~\cite{Dave-homepage} is well deserved.

\nocite{*}
\bibliographystyle{eptcs}
\bibliography{embed}

\end{document}